# Non-Invasive Near-field Spectroscopy of Single Sub-Wavelength Complementary Resonators

*Lucy L Hale[1,*], Janine Keller[2], Thomas Siday[1], Rodolfo I Hermans[1], Johannes Haase[3], John L Reno[4], Igal Brener[4], Giacomo Scalari[2,*], Jérôme Faist[2], Oleg Mitrofanov[1,4]*

*l.hale.16@ucl.ac.uk, *scalari@phys.ethz.ch

[1] Electronic and Electrical Engineering, University College London, London WC1E 7JE, UK

[2] Institute of Quantum Electronics, ETH Zürich, Zürich 8093, Switzerland

[3] Paul Scherrer Institute, Villigen 5232, Switzerland

[4] Sandia National Laboratories, Albuquerque, New Mexico

**Abstract**

Subwavelength metallic resonators provide a route to achieving strong light-matter coupling by means of tight confinement of resonant electromagnetic fields. Investigation of such resonators however often presents experimental difficulties, particularly at terahertz (THz) frequencies. A single subwavelength resonator weakly interacts with THz beams, making it difficult to probe it using far-field methods; whereas arrays of resonators exhibit inter-resonator coupling, which affects the resonator spectral signature and field confinement. Here, traditional far-field THz spectroscopy is systematically compared with aperture-type THz near-field microscopy for investigating complementary THz resonators. Whilst the far-field method proves impractical for measuring single resonators, the near-field technique gives high signal-to-noise spectral information, only achievable in the far-field with resonator arrays. At the same time, the near-field technique allows us to analyze single resonators - free from inter-resonator coupling present in arrays - without significant interaction with the near-filed probe. Furthermore, the near-field technique allows highly confined fields and surface waves to be mapped in space and time. This information gives invaluable insight into resonator spectral response in arrays. This near-field microscopy and spectroscopy method enables investigations of strong light-matter coupling at THz frequencies in the single-resonator regime.

# 1. Introduction

Near-field microscopy and spectroscopy has become an invaluable tool for studying sub-wavelength scale systems.[1,2] In particular, it enabled probing local optical fields in subwavelength-sized plasmonic resonators.[3-6] Such resonators can enhance interaction between photons and matter excitations through tight confinement of electromagnetic fields to subwavelength dimensions.[7] At terahertz (THz) frequencies, resonators were instrumental in achieving strong and ultra-strong light-matter coupling.[7]–[13] Intriguing quantum phenomena are predicted in this regime, such as spontaneous release of virtual photon pairs and unusual statistical behavior of emission from thermal sources.[14]–[16] So far, far-field techniques dominate the study of strong light-matter interaction at THz frequencies.[17] This is despite the fact that far-field scattering from single resonators is typically weak, and therefore it rarely provides sufficient sensitivity to investigate light-matter coupling in the single-resonator regime. To mitigate the weak signals, resonators have been mainly studied in arrays.[18,19] The drawback of this approach is that the inter-resonator interaction may modify the resonator spectral signature.[20 - 26] Furthermore, arrays display an average effect of many resonators and thus may limit investigations of quantum effects.[27,28] As a result, current efforts in the community point to reducing both the number of electrons coupled to the individual resonator and the overall number of resonant elements.[27]–[30] This further emphasizes the need to develop methodology for detection of weak signals from individual resonators. Although THz near-field microscopy and spectroscopy are capable of revealing characteristics of single resonators,[22], [31]–[36] so far strong light-matter interaction at THz frequencies has not been investigated in the near-field. This can be attributed to the relative complexity of near-field measurements compared to far-field techniques. In addition, potential interactions between the probe and the resonator may affect the pure resonator signature.[37]

The question remains open as to which of the techniques, far-field or near-field, is better fitted for retrieving information about subwavelength sized THz resonators with sufficient sensitivity and with minimal artefacts. To answer this, we systematically compare far-field and near-field THz time-domain spectroscopy for probing complementary metallic resonators developed for studying strong light-matter interaction in the THz frequency range.[7] We study both arrays of varying periodicity and a single resonator. We use a THz near-field microscopy technique using a collection-mode aperture probe, which has been demonstrated in applications for probing metallic and dielectric resonators,[3],[38] and a typical commercial THz time-domain spectroscopy system in the confocal configuration to detect the spectroscopic signatures. We find that both techniques are able to reveal spectral signatures of arrays, however only the near-field technique is sensitive enough to reveal spectral characteristics of a single isolated resonator. Furthermore, we observe that the inter-resonator coupling within the arrays modifies the spectral signature, making it preferable to study single resonators. The near-field technique allows us to map the electromagnetic fields both in space and time, and reveal the nature of inter-resonator coupling by visualizing surface waves travelling between the resonators on the metal-air interface. In addition, we numerically and experimentally investigate the near-field resonator-probe interaction, and find that this is negligible for probe-sample separation distances greater than 10 μm. We therefore conclude that the near-field approach is more sensitive and accurate for investigations of THz resonators in comparison to the far-field approach; it enables investigations of single resonators non-invasively, i.e. without significant effects of the resonator-probe interaction on its spectral signature. The near-field approach also has the potential for experimental investigation of inter-resonator coupling.

## 2. Experimental Section

*Complementary LC Resonators*

We study resonators with a complementary design of the split-ring LC (inductance-capacitance) resonator, with an integrated tapered dipole antenna in the center, as shown in **Figure 1**. This design is a modified version of a nanogap hybrid LC microcavity, which was used to achieve ultrastrong light-matter coupling at 300 GHz.[7] We scale the resonator such that the long axis is 50 µm, resulting in a resonance frequency of 1.1 THz, centered in the range of typical THz time-domain spectroscopy (THz-TDS) systems. The resonators are defined via electron beam lithography on a 600 µm thick GaAs substrate, metalized with a 200 nm Au film and a 4 nm Ti adhesion layer. The capacitive gap acts as a cavity of extremely reduced dimensions (1 µm) with respect to the illuminating wavelength, thus confining the electric field in a sub-wavelength sized area. Three samples are studied here: two 5x7 arrays with periodicity of 60 µm and 80 µm, and a single resonator located at the center of a gold patch.

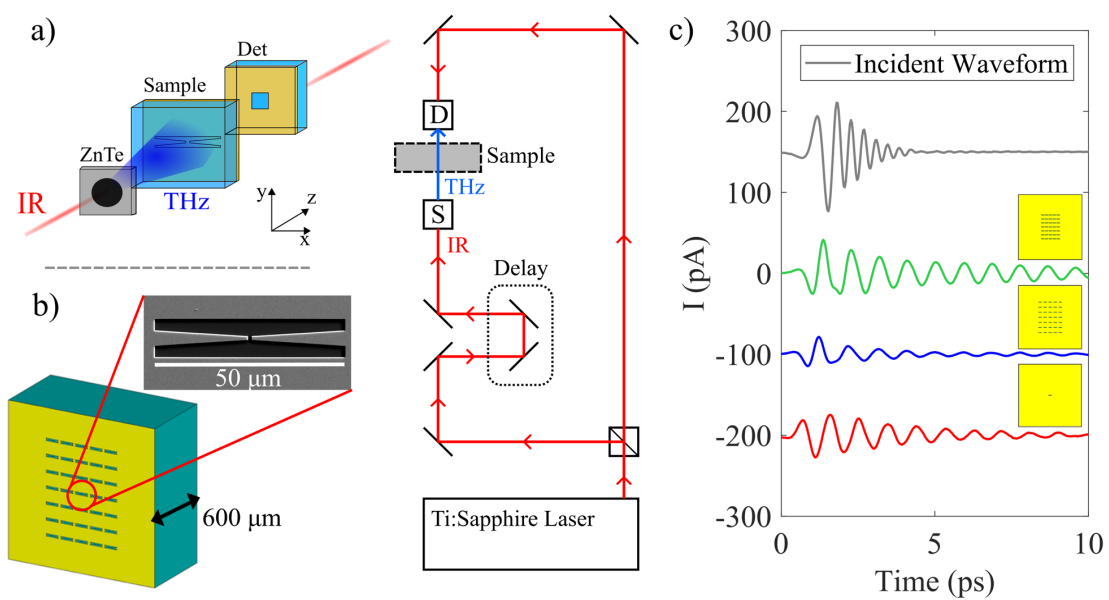

**Figure 1.** a) Schematic of transmission THz-TDS near-field system with close-up on terahertz propagation region b) Example sample (60 µm array shown) with SEM of a single array element c) Raw data waveforms for the incident waveform (grey), 60 µm array (green), 80 µm array (blue) and single resonator (red) (time traces offset for clarity)

*Near-field and Far-field Systems*

Both the near- and far-field experiments are based on transmission-type THz-TDS. Normal illumination at frequencies matching the resonator fundamental mode induce an enhanced and highly-confined electromagnetic resonance in the resonator center. Although this region is underneath the surface, evanescent fields on the resonator surface contain the information about the resonant field enhancement. These can be probed by the near-field method. At the same time, a transmitted wave also carries the information about the resonant field into the far-field. Here, we compare the spectroscopic signature of the evanescent field on the surface with that of the transmitted waves.

In both experiments, the sample is uniformly illuminated by the THz beam polarized horizontally ($x$-axis). In the near-field experiment, the illumination is set up from the substrate side, so that the THz beam passes through the GaAs substrate before reaching the resonators. In the far-field, the sample is illuminated from the metal side and the transmitted field is detected by gated antennas in the far-field (commercial THZ-TDS spectrometer, Menlo TERASMART). In the near-field approach, shown schematically in Figure 1a, a 10 μm aperture integrated with a photoconductive antenna detector [3] is placed close to the resonator side of the sample, probing the electric field in a small region near the resonator surface. The near-field probe is sensitive to two electric field components: the time derivative of the transverse component, ($dE_x/dt$), and the spatial derivative of the longitudinal component (normal to the aperture plane), ($dE_z/dx$).[39] Sensitivity to the latter component allows us to detect surface waves with a purely imaginary $k_z$-component, which do not propagate into the far-field. We note that the he detection sensitivity depends on the aperture size [40]; here we use a 10 μm aperture near-field probe [3,40], which provides an acceptable dynamic range, as can be seen in the **supporting information,** along with more detailed information about both experimental systems.

To characterize the near-field spectral signature of the resonator, we position the probe over the central region of the resonator and record a time domain waveform of the THz field. For resonator arrays, we measure the waveform of a resonator positioned in the central column of the 5-column array. We then Fourier transform the waveform and obtain the frequency-domain spectrum. After normalizing it to the incident THz pulse spectrum, measured without the sample for the same arrangement of the THz source and near-field probe, we obtain a normalized amplitude spectral density, which represents spectral enhancement $E_{det} / E_{inc}$.

*Resonator Array Spectra*

First, we use the near-field technique to probe evanescent fields for the two arrays. Their near-field THz waveforms are compared to the incident THz pulse waveform in Figure 1c. The array fields show several oscillations lasting after the decay of the incident THz pulse. Although the resonators are identical in both arrays, there are noticeable differences: the decay of the 80 μm array waveform is clearly significantly faster than that for the 60 μm array (lifetime $\tau_{80}$ = 1.7 ps in comparison to $\tau_{60}$ = 4.3 ps). This difference in the temporal field evolution indicates a different spectral response of the arrays. In **Figure 2** we illustrate this difference by comparing the normalized spectra (Figure 2a). The spectral peaks differ in linewidth: the 80 μm array is almost three times as broad as the 60 μm array in the near-field. In addition, the spectral enhancement differs – the 60 μm array enhancement is 2.25 times larger than the 80 μm array.

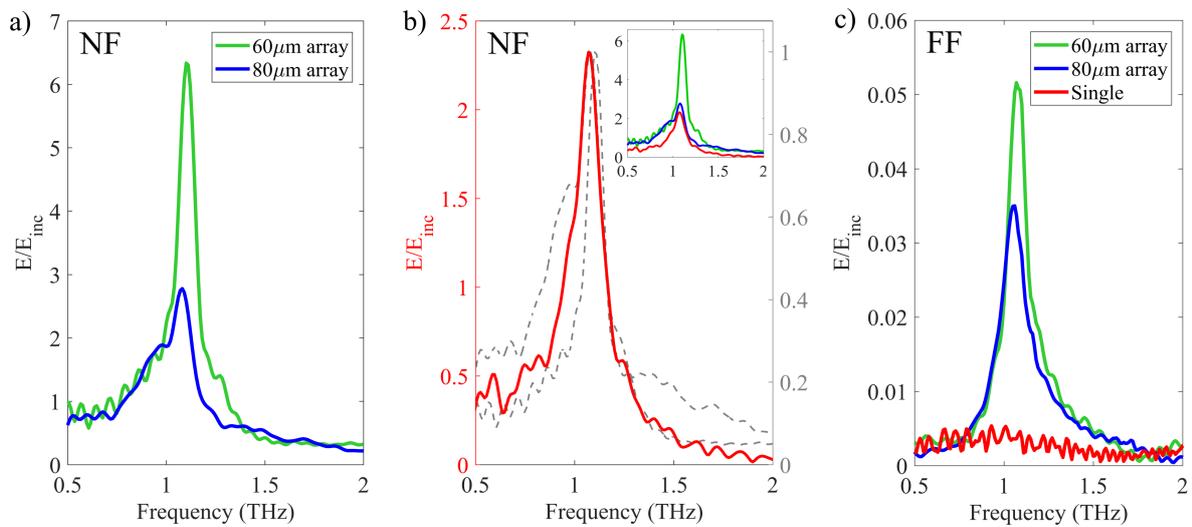

**Figure 2.** Spectral measurements: a) Near-field spectra of resonator arrays – 60 μm (green) and 80 μm (blue) b) Single resonator spectra with arrays (grey) also shown, normalized to height of single resonator spectra. Non-normalized spectra in inset. c) Far-field spectra of all three samples – 60 μm array (green), 80 μm array (blue) and single resonator (red)

We observe similar spectral differences in the far-field transmission spectra shown in Figure 2c. The linewidth is still significantly broader for the 80 μm array and the spectral enhancement is different for the two arrays. However, in the far-field this difference is smaller (linewidth is approximately 1.5 times larger for the 80 μm array and the spectral enhancement is only 1.5 times smaller). The difference in spectral amplitude from the near-field case can be attributed to the fact that in the far-field, the spectral response is affected by the array size which scales with periodicity. The peak frequencies of both arrays measured in both the near and far-field are very close to the design frequency of 1.1 THz (the spectral resolution is ≈ 60 GHz due truncation of the time-domain waveform at the first reflection within the substrate). The stark difference between the two arrays indicates that the inter-resonator coupling has a significant effect on the spectral signature of the single resonator.

*Single Resonator Spectra*

We then probe the single resonator with both systems in the same manner as the arrays. In the near-field, the single resonator has a spectral enhancement of 2.3, centered at 1.07 THz (Figure 2b). Compared to the array spectra (shown as grey lines), the single resonator spectrum is visibly different. The individual resonator linewidth can be both broader (compared to the 60 μm array case) and narrower (compared to the 80 μm case) depending on the array periodicity. In addition, we observe a variation in spectral enhancement (compared in Figure 2b, inset): for the single resonator, it is slightly lower than that for the 80 μm array (2.3 in comparison to 2.7), and almost a factor of 3 lower in comparison to the 60 μm array (6.4). Given that all three measurements were taken at similar probe-sample separations (see Section 4), this confirms that the inter-resonator coupling affects the array signatures significantly.

In the far-field, the single resonator shows no discernible spectral enhancement. This highlights a key advantage of the aperture near-field technique. In the far-field case signal-to-noise ratio is dependent on the number of resonators interacting with the beam and on radiation efficiency into the far-field. Far-field simulations, which will be discussed in Section 3.2, show that the power transmission of the single resonator is approximately 0.1% of the array transmission. This clearly demonstrates the difficulty of measuring single resonator spectra in the far-field. However, in the near-field approach the signal strength depends only on the amplitude of the field near the resonator surface, which is not significantly different between the three samples. In fact, the difference in the near-field amplitude for the same resonators measured at the same probe-sample separation can be attributed only to the inter-resonator coupling. Therefore, the near-field approach not only enables investigations at a single resonator level, it also offers quantitative evaluation of the inter-resonator interaction by detecting the field enhancement within the resonator.

## 3. Discussion

The striking result of both the near-field and far-field studies is that the spectral response from the arrays is different, despite having the same resonant element. In both experiments, the difference is evident in the spectral linewidth and field enhancement at the resonance peak. To verify that this is a result of inter-resonator coupling it is necessary to identify the physical mechanism affecting the resonator signature.

### 3.1. Inter-resonator Coupling

It was shown previously that the coupling between individual resonators in complementary arrays leads to variation of the spectral transmission signatures in the far-field.[20] When the incident beam excites the resonator, some energy scatters from the resonator in the form of surface waves propagating along the metallic surface. The surface waves are launched predominantly from resonator edges perpendicular to the beam polarization, meaning the surface wave travels across the metal plane in the direction if the incident beam polarization [39]. We note however that in our case the resonator edges responsible for launching surface waves are sub-wavelength in size and the surface waves thus diverge rapidly as they propagate from the resonator. In the arrays, the surface waves form coherent superposition at a specific wavelength, $\lambda$. This wavelength is defined by the array periodicity, and is given by the interference condition, which arises from the dispersion relation of surface plasmon waves at an interface:

$$\lambda = a\sqrt{\varepsilon_{eff}}\left(\sqrt{i^2 + j^2}\right)^{-1} \tag{1}$$

where $i, j$ are integers, denoting the order of the reciprocal lattice vector, $a$ is the lattice constant of the array, and $\varepsilon_{eff}$ is the effective relative permittivity of the structure. Previous numerical simulations of infinite arrays fulfilling the interference condition (Equation 1) showed a splitting of the spectral peak (not shown here, more detail in the supporting information). Such splitting was interpreted as a strong coupling between the resonator mode and the surface wave mode, and the effective relative permittivity was found to be: $\varepsilon_{eff}$ = 11.6.[20]

Using this model in the present experiment, we expect the lowest order surface wave mode at an array periodicity of 80 μm at the GaAs-Gold interface to be at a frequency of ~1.1 THz ($i$=1, $j$=0). This frequency coincides closely with the resonance frequency of the single array element. Therefore, in addition to the resonant excitation of each element by the incident pulse, each element is also excited by the surface waves. As the periodicity of the array matches the wavelength of the surface wave mode, the array elements are excited in phase. This produces a collective interference effect, affecting radiative losses of the resonator array. The interference causes an increase in spectral linewidth, which corresponds to an increase of radiative losses and decrease in quality factor. On the other hand, for the 60 μm array, the surface wave mode wavelength is shorter than the period and thus the secondary excitation of the array elements does not occur in phase with the resonance excitation. Interestingly, this results in a narrowing of the observed resonance corresponding to a reduction of radiative losses for the 60 μm array - opposing the superradiance effect, which results in a broadened linewidth with reduced resonator spacing.[41]

### 3.2 Finite Array Simulations

Experimental measurements do not feature the splitting predicted by infinite array simulations of constructively interfering resonators. This suggests that ideal infinite simulations provide a qualitative description of the physical coupling mechanism, but do not accurately describe the experiment. To obtain an improved description, we carry out simulations of the entire 5x7 array, without using periodic boundary conditions. The results of this are shown in **Figure 3**.

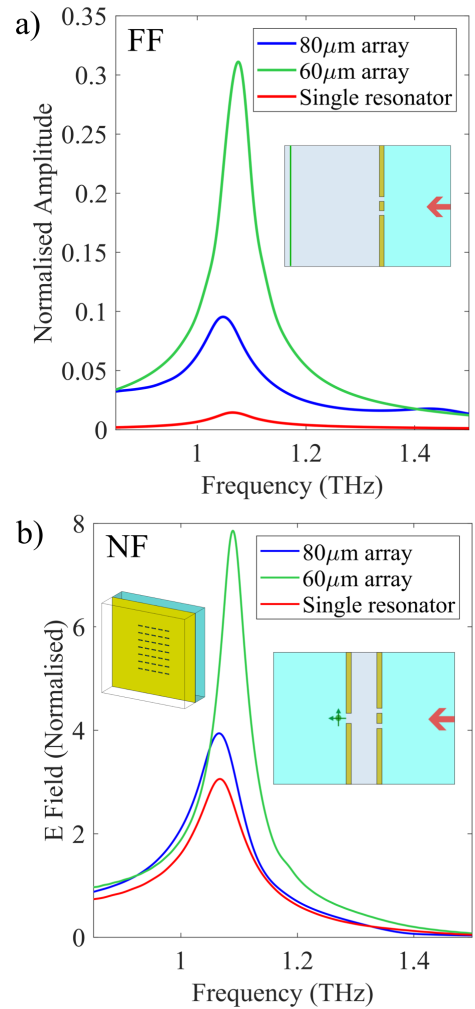

**Figure 3.** Simulation results: a) Simulated spectra in the far-field (interpolated) - measured transmission through face on the far end of the simulation region (shown in inset) b) Simulated near-field spectra measured using probe placed 1μm behind aperture (simulation region shown in inset).

Initially, we simulate the far-field case (Figure 3a). The samples are illuminated through the substrate, and the power transmission through the far side of the simulation region is calculated. This is illustrated by insets in Figure 3a, and further details of the specific simulation set-up can be found in the supporting information. As with the experiment, we see no splitting of the resonance peaks. The simulated spectra are in good agreement with the experiment: the 60 μm spectrum is significantly higher in amplitude, and it has a smaller linewidth than the 80 μm array.

Furthermore, we numerically model the near-field detection by introducing the aperture probe into the simulation region (Figure 3b). The aperture plane is placed 10 μm away from the resonator plane (estimated as a lower bound for the probe-sample separation in these experiments, as discussed in Section 4). The structure is illuminated through the substrate and the electric field is detected behind the aperture. As with the far-field simulation, no spectral peak splitting is observed. The spectra match the near-field experiment well: the linewidth of the 80 μm array (0.12 THz) is much larger than for the 60 μm array case (0.07 THz); and the decreased spectral enhancement for the 80 μm array is also seen. The relative spectral enhancements of the three simulated structures match well to the experimental data in Figure 2.

The agreement of the experimental results with simulations of the entire structure confirms that the lack of peak splitting and the varying spectral enhancement observed experimentally are genuine consequences of the finite nature of the array, rather than experimental error or lack of spectral resolution.

In order to explain the difference between the infinite and finite array simulations, we consider the surface wave induced inter-resonator coupling. For the finite array simulation, we find that the strongest field is localized at the resonators, as expected. However, we also observe fields on the gold surface as far as several hundred microns away from the array (shown in supporting information Figure S2). This shows that surface waves are excited at each resonator and travel along the metallic surface, exciting neighboring resonators as they travel. When there is a phase difference between the initial excitation of the resonator and the surface wave excitation, interference occurs, which varies for each resonator depending on its position in the finite array. The infinite array model however fails to describe the field distribution in the finite array. Since each element of the finite array experiences different secondary excitation from the surface wave, both in amplitude and phase, the finite array spectrum is naturally 'blurred' in comparison to the infinite array.

A more detailed description of the physical mechanism for this is provided in the supporting information.

### 3.3 Near-Field Observation of Surface Waves

Surface waves can be observed by the near-field system. This provides further insight into resonator operation and inter-resonator interactions. Although the near-field technique cannot confirm the presence of the surface wave on the gold-GaAs interface, it does allow us to observe surface waves on the gold-Air interface. **Figure 4** shows an *x*-axis line scan (red line in the scanning electron microscope (SEM) image) covering three resonators in the 80 μm array. The area scanned spans from the central resonator to the edge resonator, and extending about 300 μm beyond the array edge. The near-field space-time map shows that the surface waves travel over 200 μm from their source whilst still maintaining a detectable amplitude. This therefore confirms that surface waves are launched and supported by the structure, and that they have sufficient energy to excite neighboring resonators. We can directly measure the effective permittivity from the space-time gradient, and therefore wavelength of the surface wave on the gold-air interface from Figure 4. Unlike the surface waves on the gold-GaAs interface, it has the *k*-vector closer to the free space vector and therefore a period of 223 μm for a 1.1 THz wave (corresponding $k = 1.22k_0$). Given that the gold layer (200 nm) is comparable to the penetration depth, the surface waves on both interfaces will extend through the metal layer . However, the gold-air surface wavelength is several times longer than the array periodicity, confirming that it is the surface waves on the gold-GaAs interface that contribute to coupling, not the surface waves on the gold-air interface, which we observe here in near-field maps.

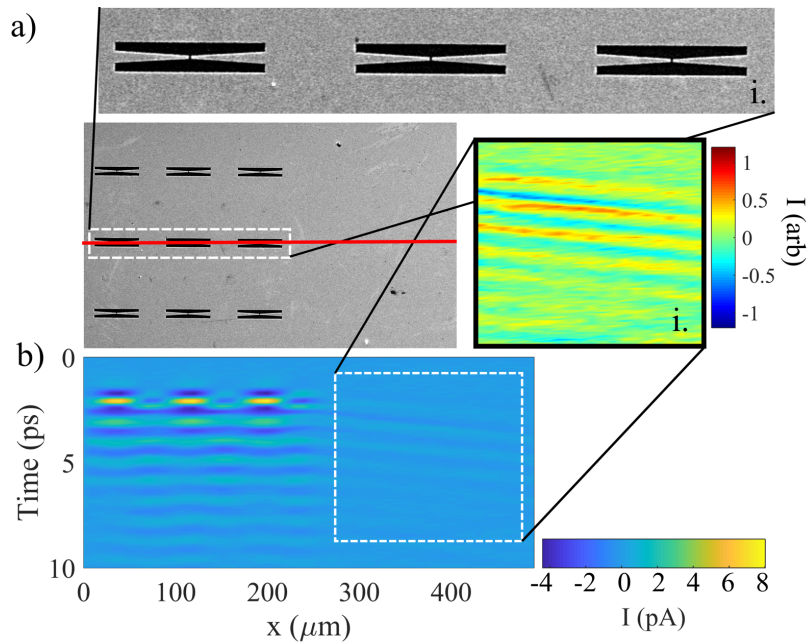

**Figure 4.** Space-time Map: a) SEM of three resonators in 80 μm array scanned in space-time map. i) Larger SEM of 80 μm array, indicating line scanned in space time map. b) Experimental near-field space-time map spanning a line of the 80μm array (shown by red line in a). i)) Rescaled close up of space-time map illustrating surface waves travelling in time on the GaAs-Gold interface.

The effect of the surface waves is evident in the detected fields at the resonator center: we observe a different time evolution of the field for resonators in different positions in the array. Although all resonant elements are excited in phase by the initial THz pulse (the beam spot size is ~1-2 mm, which is much larger than the distance between the resonators), the secondary excitation due to the surface wave varies for different resonators, resulting in a phase difference between the resonant fields at neighboring elements. This agrees with the finite array simulated results (provided in **supporting information**). The non-zero phase difference between different resonators within the array also allows us to distinguishes these surface waves from magnetoinductive waves, which have been detected in similar systems. [42,43].

We also observe the effect of surface waves in the spatial distribution of the experimental near-field maps for a single resonator and an array. **Figure 5** compares near-field spatial maps of one of the resonators in the 80 μm array (Figure 5a), to the single isolated resonator (Figure 5c). Both were recorded at a time corresponding to the maximum of the waveform. In both images the resonator shows enhanced fields in comparison to the surrounding area. However, for the resonator in the array, whilst the individual array elements can be identified, there is considerable field intensity in between the resonators. This is consistent with surface waves on the metal-air interface.

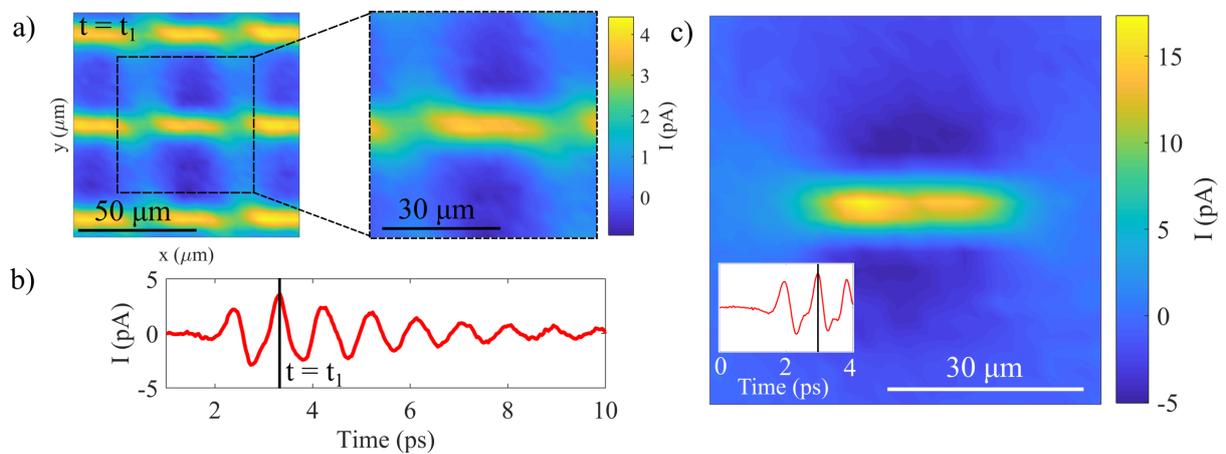

**Figure 5.** a) Interpolated near-field map of 80μm array  b) Waveform recorded from 80 μm array showing time delay which the near-field map in (a) was recorded at, corresponding to the maximum amplitudes. c) Interpolated near-field map of single resonator at time corresponding to maximum of the waveform (inset).

As a result of simulations we can conclude that our experimental observations of varying linewidth and spectral enhancement are a direct result of inter-resonator coupling due to surface waves on the gold-GaAs interface. The lack of spectral peak splitting is a consequence of the finite nature of the array, showing that the array is too small to gain quantitative information from infinite simulations. Whilst we cannot experimentally observe the field on the gold-GaAs interface, the near-field technique allows us to detect surface waves on the gold-air interface in near-field spatial and space-time maps. The measured wavelength implies gold-air surface waves do not contribute significantly resonator coupling. However, their observation strongly suggests that they are present on the GaAs interface, and this results in the varying spectral characteristics of the samples.

## 4. Probe-Sample Interaction

We now address the question of probe-sample interaction. The near-field probe positioned within the mode of the resonator can affect its spectral characteristics, and therefore limit or invalidate the useful capabilities of near-field methods [33]. The effect of the probe on spectral characteristics depends on its proximity to the resonator, and the probe-sample interaction can be evaluated by quantifying the distance dependence of the main characteristics of the resonator: the resonant frequency and the linewidth. Since the aperture-type method discussed here can be applied at any distance, we experimentally record spectra of the resonator for various separations between the probe and the sample. We also numerically simulate this by modelling the full resonator structure and aperture probe (as described in Section 3.2), and measuring the spectra through the aperture for various distances between the resonator and aperture plane. The measured spectra are shown in **Figure 6**.

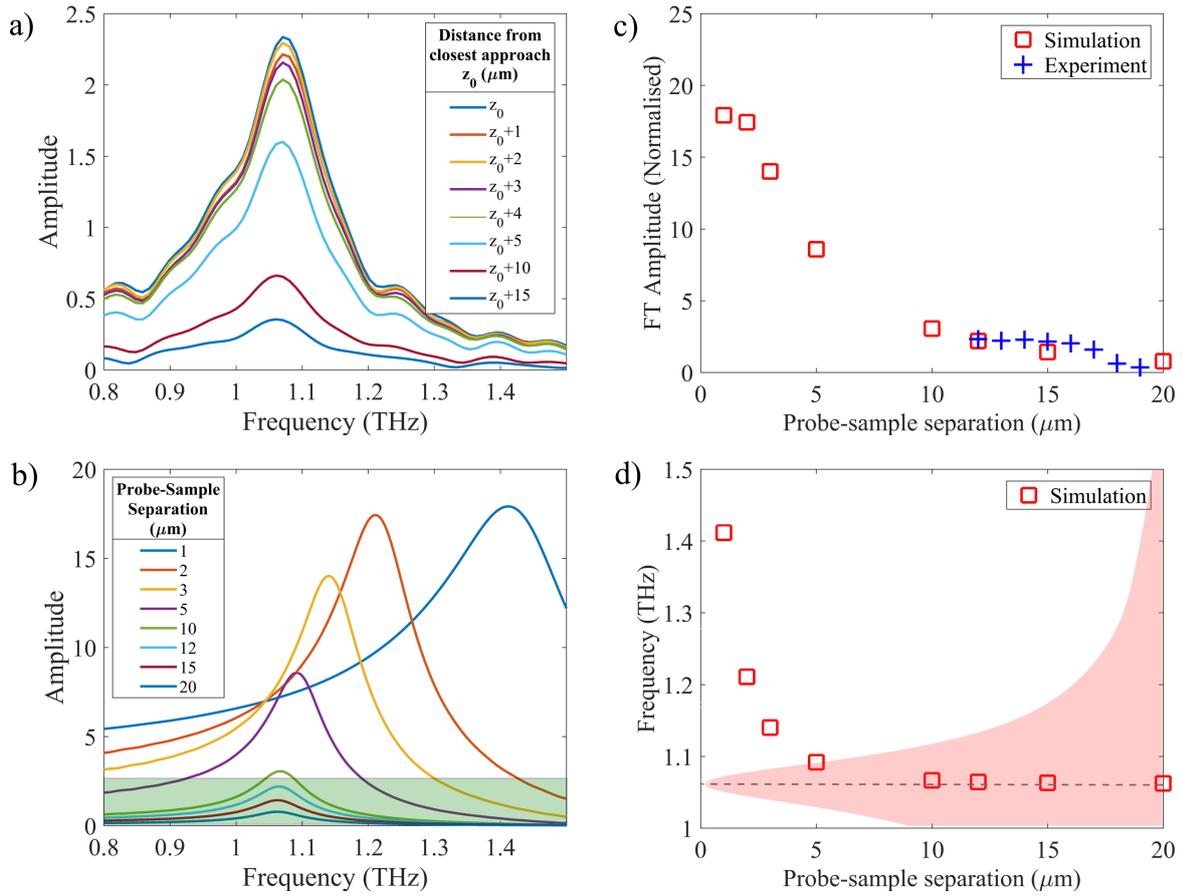

**Figure 6.** a) Experimentally measured spectra of single resonator for increasing probe-sample separation distance ($Z_0$ – position of closest approach). b) Simulated spectra of single resonator for increasing probe sample separation (1μm - 20μm). c) Peak Fourier transform amplitude (from b) is plotted against the probe-sample separation for simulation, with experimental data (from a) fitted. The experimental probe-sample separation for the closest approach can be estimated to be approximately 10-15 μm. d) Simulated frequency position of Fourier transform peak for increasing probe-sample separation with simulated peak shown (faded red) to illustrate linewidth.

We consider the aperture of the near-field probe positioned along the optical axis that passes through the region of strong field confinement at the resonator center. As we move the probe further from the resonator, the amplitude of the resonant peak decreases - it practically disappears at the probe-sample distance of ~ 20 μm. However, within a range of probe-sample distances (~10-20 μm), we don't observe a detectable change in the resonance frequency or in the linewidth (Figure 6a). Using numerical simulations, we extend the range of studies to smaller distances (Figure 6b), where we find that the frequency of the spectral peak starts increasing for sample-probe separations smaller than ~10 μm, with a sharp increase for distances smaller than 5 μm. Figure 6c and 6d illustrate the frequency and amplitude change with probe-sample separation using the full array simulations. The simulations confirm that at distances greater than 10 μm there is very little effect of the probe on the frequency of the resonator, whereas below ~10 μm, the probe starts increasingly altering the spectral properties of the sample. The amplitude of the resonance also strongly depends on the distance. The amplitude dependence allows us to estimate the probe-sample separation in the experiment by fitting the experimental amplitude dependence to simulation data (blue crosses in Figure 6c). From this process we verify that the spectra presented in Figure 2 correspond to the probe-sample separation of approximately 10-15 μm. At this distance from the sample surface, the near-field signal from the resonator is strong enough to be detected by our probes, whereas there is a negligible effect of the probe on the spectral signature of the resonator. It is worth noting that whilst this is the case for the specific resonator structures here, the distance between the probe and sample at which the effect of the probe is negligible will vary depending on the resonator structure and field confinement.

At a distance of 10 μm, the probe is relatively far away from the region of strong field confinement, which is determined by the antenna gap in the resonator (in our case is ~1 μm). Nevertheless, the spectral signature of the resonator manifests itself in the spectrum of the surface waves propagating along the antenna arms. In Figure 5b we find areas of highest measured field amplitude along the antenna arms, rather than at the resonator center. This is because the probe is far away from the confined field region in the gap to detect it directly. At the wider section of the antenna however, the antenna field is not as strongly confined as at the center, and therefore it extends further from the resonator surface, giving the appearance of a stronger field magnitude at the antenna arms. Therefore, the probe positioned at 10 μm from the resonator allows us to analyze the resonant fields without introducing the probe in the region of strong field confinement in the antenna gap. The near-field approach therefore can be applied for spectroscopic analysis of resonators with substantially stronger confinement.[7], [13]

## 5. Conclusion

In conclusion, we investigated complementary THz resonators in arrays and as a single isolated object using far-field transmission-mode THz spectroscopy and collection-mode near-field THz spectroscopy. Whilst both methods are capable of retrieving spectral signatures from resonator arrays, only the near-field method enables high signal-to-noise spectral measurements of individual resonators. Furthermore, near-field imaging provides even deeper insight into resonator operation in arrays. Infinite array models fail to predict array spectra correctly. Finite array simulations however, accurately predict the spectra measured both in the near- and far-field systems. Moreover, we can experimentally detect surface waves on the Metal-Air interface using near-field imaging, and rule out their contribution to inter-resonator coupling. We therefore confirm that surface waves on the GaAs-Metal interface cause inter-resonator coupling which modifies the spectral signature. Finally, we show that the effect of the near-field probe on the spectral properties of this specific resonator can be eliminated by maintaining a probe-sample separation of ~10μm.

This work demonstrates the potential of collection-mode near-field THz spectroscopy and imaging for studying strong light-matter interactions with high signal-to-noise, and without influence of interference from neighboring resonators or the near-field probe. As the detection is not determined by far-field scattering efficiency, even studies employing extremely confined fields to couple to very few and even single electronic oscillators could be done in the near-field. We anticipate that the near-field approach implemented in more complex experiments with cryogenic environments and DC magnetic fields will be integral to exploring strong and ultrastrong coupling phenomena at THz frequencies.

## Acknowledgements

LH acknowledges financial support from EPSRC, project EP/L015455/1 "Centre for Doctoral Training in Integrated Photonic and Electronic Systems' and TS acknowledges financial support from EPSRC, project EP/L015277/1 "Centre for Doctoral Training in the Advanced Characterisation of Materials". RH and OM acknowledge financial support from EPSRC, project EP/P021859/1 "HyperTerahertz". JK, JH, JLR, IB, GS, JF acknowledge financial support from the Swiss National Science Foundation (SNF) through the National Centre of Competence in Research Quantum Science and Technology (NCCR QSIT) and Molecular Ultrafast Science and Technology (NCCR MUST). JK, JH, JLR, IB, GS, JF acknowledge financial support from the ERC grant MUSiC. We thank Felice Appugliese for help with some far field measurements. This work was supported by the U.S. Department of Energy, Office of Basic Energy Sciences, Division of Materials Sciences and Engineering. Fabrication of THz near-field probes were performed at the Center for Integrated Nanotechnologies, an Office of Science User Facility operated for the U.S. Department of Energy (DOE) Office of Science. Sandia National Laboratories is a multimission laboratory managed and operated by National Technology and Engineering Solutions of Sandia, LLC., a wholly owned subsidiary of Honeywell International, Inc., for the U.S. Department of Energy's National Nuclear Security Administration under contract DE-NA-0003525. This paper describes objective technical results and analysis. Any subjective views or opinions that might be expressed in the paper do not necessarily represent the views of the U.S. Department of Energy or the United States Government.

# Supporting Information

**Non-invasive Near-field Spectroscopy of Single Sub-Wavelength Complementary Resonators**

*Lucy L Hale[1,*], Janine Keller[2], Thomas Siday[1], Rodolfo I Hermans[1], Johannes Haase[3], John L Reno[4], Igal Brener[4], Giacomo Scalari[2,*], Jérôme Faist[2], Oleg Mitrofanov[1,4]*

## 1. Experimental Set-Ups

*Far-Field:* A commercial Menlo TERASMART system is used for far field measurements. An ultrafast fiber coupled femtosecond laser operating at 1560 nm generates a broadband THz pulse. THz antenna receivers and emitters used for source and detection (TERA15-FC).

*Near-Field:* Broadband THz pulses are generated through optical rectification in a ZnTe crystal, using 100 fs pulses from a Ti:Sapphire laser. Detection is done using an aperture integrated photoconductive antenna (10 μm in size) that is placed 10-20 μm away from the sample surface. **Figure S1** shows the dynamic range of this set-up.

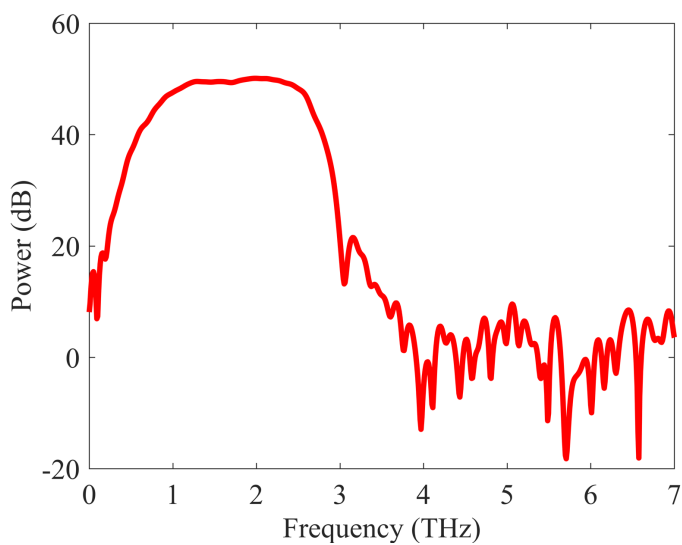

Figure S1: The dynamic range of the system with a 10 μm aperture probe, calculated from the Fourier transform when no sample present

## 2. Aperture Probe Details

The position of the aperture probe over the sample determines what field components are collected. When the aperture is centered over the resonator at sufficiently close distance the signal is dominated by the $dE_x/dt$ component. However, when the aperture is positioned over the metallic surface, only surface waves are present, and the detected field is purely evanescent and is polarized normal to the surface.

In the spectroscopy measurements in this work, the aperture was placed over the central region of the resonator, at a distance >10 μm away from the resonator plane. As a result, the detected field is a combination of the $dE_x/dt$ component from the central region, and the $dE_z/dx$ components dominated by the surface waves from the surrounding metallic surfaces.

## 3. Calculating Surface Waves

The dispersion relation of surface plasmon waves on an interface given by:

$$k = \frac{\omega}{c} \sqrt{\frac{\varepsilon_1 \varepsilon_2}{\varepsilon_1 + \varepsilon_2}}$$

Where $\varepsilon_1$ and $\varepsilon_2$ are the permitivitties of the two materials at the interface. We simplify the equation and use an effective permittivity in order to consider the finite thickness of the metal and the fact that at THz wavelengths, the real part of the metal permittivity is negative and order of magnitudes higher than the dielectric either side:

$$k = \frac{\omega}{c} \sqrt{\varepsilon_{eff}}$$

From this we arrive at equation (1) with which we find the surface wave wavelength. To calculate the effective permittivity, we follow the method outlined in [20]. We simulate a unit cell of the infinite structure with periodic boundaries, for varying periodicity. The frequency of spectral peaks is used to determine the coupling strength between the resonator mode and the surface wave mode. From this we find an effective permittivity of $\varepsilon = 11.6$.

## 4. Finite Array Simulations

The full finite 5x7 array (or single resonator) is simulated on a 1000 μm x 1000 μm gold sheet (thickness 300 nm) using an electrical conductivity of 4.561 x $10^7$ S/m, on a GaAs substrate (thickness 200 μm) in a vacuum background, with open (PML) boundaries on all sides. A plane wave illuminates the structure from behind the resonator plane (as indicated by the red arrow on Figure 3a,b inset).

*Far Field Simulations:* The power transmission is calculated through the face parallel to the resonator plane on the far side of the simulation region (200 μm away from resonator plane).

*Near Field Simulations:* The aperture plane is modelled by a gold sheet of the same size and thickness as the resonator plane, with a 10 μm square aperture in the center (so that it is positioned over the central resonator), and GaAs behind to model the photoconductive layer. The electric field is detected by a field probe placed 1 μm behind the aperture (indicated in Figure 3a inset by green marker). The corresponding electric field spectra are normalized to the spectrum obtained for the case when the resonator is absent.

*Finite Array Field Distribution:* To investigate surface waves at the GaAs-Gold interfaces in the finite array, we use numerical simulations, and in **Figure S2** we display the maximum time-domain electric field map of the array. We observe the strongest field is localized at the resonator centers, and we also observe fields on the gold surface travelling several hundred microns away from the array. As shown in Figure S2, the field distribution for different resonators in the array is not identical, in contrast to the infinite array, where periodic boundary conditions are imposed, forcing identical behavior of all resonators in an infinite array. For example, the column of resonators at the finite array center displays a field minimum on the metal plane exactly in line with the resonator center, whereas a similar minimum shifts in the neighboring resonators, because the periodicity is not matched perfectly to the surface wave k-vector. As a result, a resonator in the next column, with one neighbor to the left and three neighbors to the right has an asymmetric field distribution at its surface. The asymmetry becomes even more obvious for the resonators at the array edge. This shift in the position of the minimum translates into different primary and secondary waves at each resonator. Since each element of the finite array experiences different excitation conditions, the finite array spectrum is altered from the infinite array.

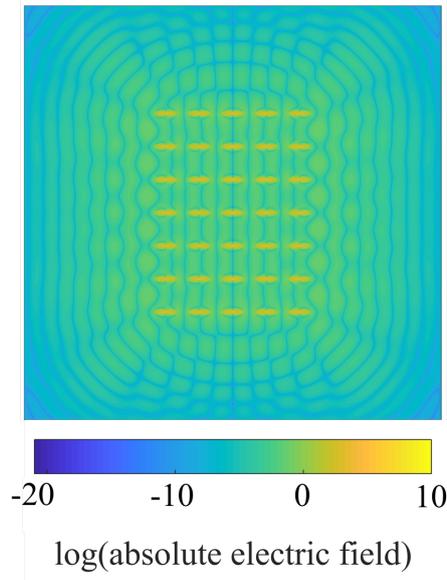

Figure S2: Simulated absolute electric field (logarithmic scale) on GaAs-Gold interface of 80 μm array at a snapshot in time (maximum field) illustrating surface waves.

## 5. Discussion of Aperture Size and Spatial Resolution

We consider the size of the aperture in the near-field approach. The aperture size allows us to find an acceptable trade-off between resolution and signal-to-noise ratio. Whilst the 10 μm aperture used here is too large to resolve the central cavity of the resonator, or to give fine details of the field distribution, it gives a high signal-to-noise level (~$10^2$ in field amplitude, determined experimentally using the illumination field without a sample). At the same time, the 10 μm aperture is much smaller than the wavelength, and therefore the transmission properties of the aperture can be described by a smooth analytical function of frequency $E_t$ /

$E_{inc} \sim \omega$ and the aperture does not introduce resonant features in the measured spectra. Change of transmission with aperture size is outlined in [40]. A smaller aperture would result in higher spatial resolution, however at the expense of signal-to-noise ratio, as the detected signal decreases with the cube of aperture size. Furthermore, the spatial resolution is also limited by the probe-sample separation. Since, it is preferable to use a probe-sample separation of >10 µm in order to avoid the interaction of the probe with the resonator, a smaller aperture would not improve the spatial resolution, as it will not isolate the field at the resonator center. Therefore, we find that a relatively large aperture (~10 µm) works better for practical spectroscopy of complementary resonators, as it allows us to detect the spectral properties of the resonator via evanescent fields on the resonator surface, whilst maintaining a high signal-to-noise.